\documentclass[%
reprint,
amsmath,amssymb,
prstab,
]{revtex4-1}

\usepackage{graphicx}
\usepackage{float}
\usepackage{dcolumn}
\usepackage{bm}
\usepackage[colorlinks,allcolors=blue]{hyperref}

\begin{document}

\preprint{APS/123-QED}

\title{Coherent X-rays with Tunable Time-Dependent Polarization}

\author{N. Sudar, R. Coffee, E. Hemsing}
\affiliation{SLAC National Accelerator Laboratory, Menlo Park, California 94025, USA
}%

\date{\today}

\begin{abstract}
We describe a method for producing high power, coherent x-ray pulses from a free electron laser with femtosecond scale periodic temporal modulation of the polarization vector.  This approach relies on the generation of a temporal intensity modulation after self seeding either by modulating the seed intensity or the beam current.  After generating a coherent temporally modulated $s$-polarization pulse, the electron beam is delayed by half a modulation period and sent into a short orthogonally oriented undulator, serving as a $p$-polarization afterburner.  We provide simulations of three configurations for realizing this polarization switching, namely, enhanced self seeding with an intensity modulation generated by 2 color self seeding, enhanced self seeding of a current modulated bunch, and regular self seeding of a current modulated bunch.  Start to end simulations for the Linac Coherent Light Source-II are provided for the latter.
\end{abstract}

\maketitle
\section{Introduction}
The x-ray Free Electron Laser (FEL) has provided the scientific community with a tunable source of coherent, high peak power x-rays capable of probing phenomena at the atomic scale \cite{RevModPhys.88.015006}.  Over the past decade, the scientific capabilities of FEL facilities have been expanded by increases in the peak brightness and repetition rate \cite{FELS2017,EuroXFEL}, further control of the spectral and temporal properties \cite{ATTO,PhysRevLett.110.134801}, and improvements in the stability of FEL pulses \cite{Giannessi2020}.  In anticipation of the emerging needs of the user community, investigation of novel configurations aimed at tailoring the spectral, spatial, temporal, and polarization properties of FEL photons is an active area of research.

Recently, several methods have been proposed to generate x-ray pulses with structured spatial and temporal polarization topology.  These include the generation of x-ray beams with few-$\mu$m scale spatially varying polarization and 100 fs scale temporally varying polarization \cite{PhysRevAccelBeams.22.110705}, the generation of Poincar\'e beams \cite{Morgan_2020}, and the generation of attosecond scale temporally varying polarization \cite{morgan2020attosecond}.  Each of these approaches share a common theme, using an FEL afterburner consisting of crossed undulators to generate and overlap radiation with orthogonal polarization or opposite helicity.  This purely electron beam-based approach removes the need for polarizing optics not readily available at x-ray wavelengths.

One of the reasons for the growing interest in polarization control \cite{Gregor2016} stems from the desire to probe electronic symmetry as it relates to induced circular dichroism in atomic systems \cite{Tomasso2016,MarkusPRL2017,Ilchen2018} and chirality in molecular systems \cite{Markus2017,Gregor2019}.  The high degree of symmetry in atomic systems enforces the familiarly strict transition selection rules that dominate the photo-excitation process in the optical regime.
In the x-ray regime, however, transition energies lie far above valence ionization thresholds such that the electronic symmetries involved also couple to the outgoing free electron partial wave.  Induced chirality can occur when these selection rules become modified by dressing the atomic ensemble with an optical field.  In such cases, the various symmetries of resonant core-to-valence transitions produce lab-frame measurable modulations to photo-electron spectra.  Access to the potentially time-evolving electronic symmetries in molecular systems as they explore internal degrees of freedom motivates targeted femtosecond (fs) scale polarization control at x-ray FELs.

Natural linewidths in the soft x-ray regime are in the order of a few hundred meV.  One therefore requires a few fs for a coherence to develop between the x-ray field and the molecular system.  Resonant excitation is the result of this developing coherence.  Now, if the driving x-ray field switches polarization state, one could effectively modulate the core-to-valence resonant transition at an optical frequency while allowing for many cycles of x-ray--molecule coherence to develop.  Following a few optical cycles of such a dithered excitation, one would expect the preparation of an optically active coherent ensemble of mixed electronic nature.  

Given the rising interest in chirality in molecular systems, here we propose a path toward tunable fs-scale polarization switching in xFELs that may allow examination of these systems.  First, a longitudinally coherent, temporally modulated intensity profile is generated in the $s$-polarization.  
This intensity modulation will impart a periodic modulation on the electron beam bunching profile.  
Introducing a magnetic chicane delay, the electrons are delayed by approximately half the intensity modulation period.  
Provided this delay is not too large, the electron bunching is preserved, or potentially even enhanced by the optical-klystron effect \cite{PhysRevLett.96.224801}.  
Sending the electrons through an orthogonally oriented undulator section, an FEL pulse with nearly identical intensity profile is generated in the $p$-polarization, shifted in time relative to the $s$-polarization pulse by half a modulation period.  The preservation of the initial bunching between stages gives a stable phase relationship between the two pulses.  This allows for fine adjustment of the phase shift between pulses, potentially producing a smooth temporal rotation of the polarization vector across the pulse.  This scheme can be applied to the generation of time-varying circular polarization using a series of helical undulators of opposite helicity as well. 

The main challenge presented by this method is the generation of the required fs-level intensity modulation.  Herein we describe three possible configurations, each with unique benefits.  The first uses the enhanced self seeding (ESS) method described in \cite{PhysRevLett.125.044801}, using the two-color self seeding technique to generate a beatwave intensity modulation which is subsequently amplified by a fresh electron bunch with a flat current profile.  The second again uses the ESS method, this time generating a monochromatic seed that is amplified by a fresh electron bunch with a modulated current profile.  The third uses regular self seeding (SS) with a current modulated electron bunch as described in \cite{PhysRevSTAB.15.050707}.      

In this paper, we first provide a simplified model of the generation of time-varying polarization with a brief overview of the useful Stokes vector formalism.  In section III, we discuss the three methods for generating the requisite intensity modulation in greater detail.  This includes simulations with idealized conditions and uses undulator and electron beam parameters relevant to the Linac Coherent Light Source-II (LCLS-II), using the 3-D, time dependent FEL code, {\sc genesis} \cite{REICHE1999243}.  Section IV describes a potential method for the generation of a fs-level current modulation at LCLS-II, including simulations of the LCLS-II linac.  Included are full start to end simulation results using the method described in section III. 

\section{Time Varying Polarization: Simple model}

Our method to produce time-dependent polarization relies on the combination of two orthogonally polarized pulses with identical temporal structure, shifted in time. Consider an electric field produced by the emission of radiation from an electron beam propagating through two consecutive undulator sections. In the first undulator, the polarization of the emitted radiation is $\mathbf{\hat e}_1$. The electron beam then passes through a delay of $\Delta t$, and then through another undulator where it radiates an identical pulse, but with polarization $\mathbf{\hat e}_2$. The total electric field vector can be written as
\begin{equation}\label{totfield}
\mathbf{E}(t)=E_0(t+\Delta t/2)\mathbf{\hat e}_1+e^{i\psi}E_0(t-\Delta t/2)\mathbf{\hat e}_2=\begin{pmatrix} 
E_x \\
E_y\\
\end{pmatrix}.
\end{equation}
Here $\psi$ is a small phase shift between the two pulses. 

To describe the polarization properties of the radiation, it is useful to examine the individual components of the Stokes vector $\mathbf{S}=(S_0,S_1,S_2,S_3)$;
\begin{equation}
\begin{aligned}
S_0&=|E_x|^2+|E_y|^2\\
S_1&=|E_x|^2-|E_y|^2\\
S_2&=E_xE_y^*+E_x^*E_y\\
S_3&=i(E_xE_y^*-E_x^*E_y)
\end{aligned}
\end{equation}
The first Stokes parameter $S_0$ is the temporal intensity profile. The $S_1$ parameter describes linear polarization in either the horizontal or vertical direction, depending on the sign. $S_2$ describes diagonal linear polarization in the $\pm45^\circ$ directions, and $S_3$ describes circular polarization.

To produce time-dependent polarization in Eq.~(\ref{totfield}), we consider each polarization component to be described by a field $E_0(t)$ with carrier frequency $\omega_0$ and with a periodic temporal intensity modulation,
\begin{equation}\label{E0}
E_0(t)=e^{i\omega_0t}\cos(\Delta\omega\, t/2)=\frac{e^{i\omega_1t}+e^{i\omega_2t}}{2}.
\end{equation}
This is mathematically equivalent to a pulse comprised of two frequencies separated by $\Delta\omega=\omega_1-\omega_2$ as would be radiated by an electron beam coherently bunched at these two frequencies. Time-dependent polarization in Eq.~(\ref{totfield}) comes from shifting the differently polarized pulses in time by $\Delta t$ to interleave the intensity modulations, and by finely adjusting their relative phase with $\psi$. 
 
\begin{figure}
       \includegraphics*[width=0.99\linewidth]{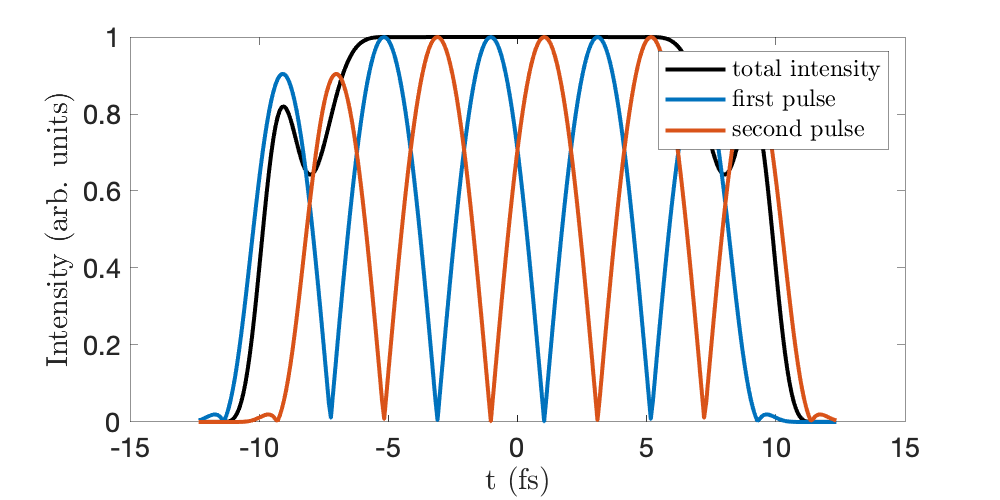}
   \caption{Example pulse intensity profiles from consecutive crossed undulators. Each pulse carries two colors separated by $\hbar\Delta\omega=1$~eV at $\hbar\omega_0=1.24$~keV. The pulses from each undulator are interleaved with a time delay of $\Delta t=\pi/\Delta\omega=2$~fs.}
   \label{fig:pulses}
\end{figure}

Let us consider the case of two orthogonally polarized planar undulators. The field polarization unit vectors are 
\begin{equation}\label{vecs}
\mathbf{\hat e}_1=\mathbf{\hat e}_x=\begin{pmatrix} 
1 \\
0
\end{pmatrix},\qquad \mathbf{\hat e}_2=\mathbf{\hat e}_y=\begin{pmatrix} 
0 \\
1
\end{pmatrix}
\end{equation}

\begin{figure}[h]
       \includegraphics*[width=0.99\linewidth]{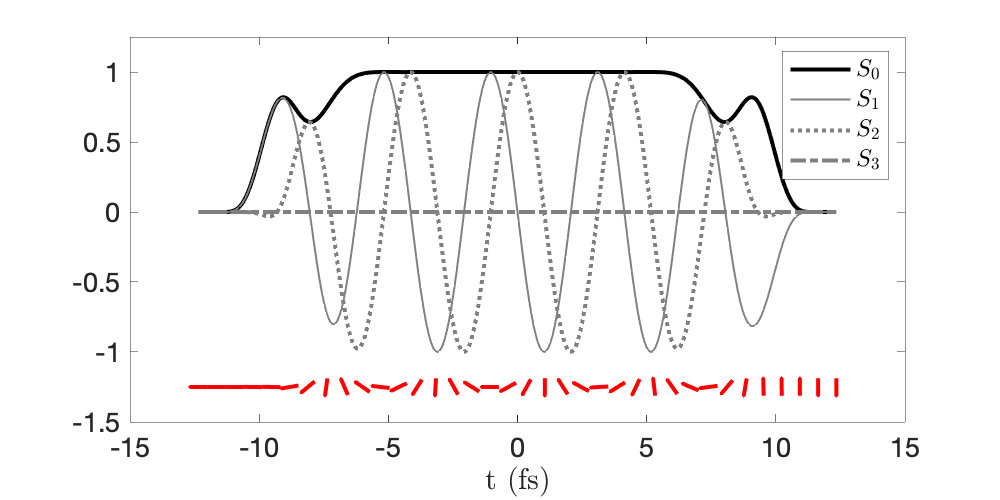}
        \includegraphics*[width=0.99\linewidth]{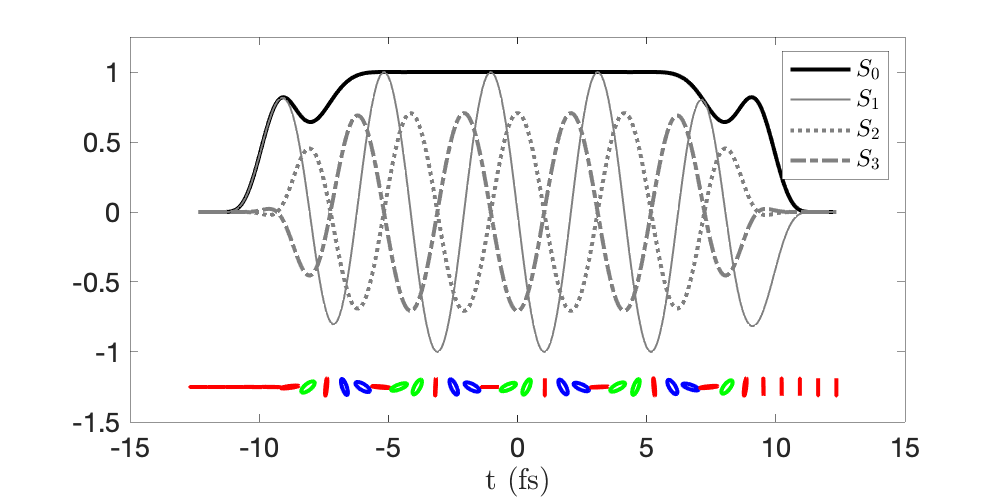}
         \includegraphics*[width=0.99\linewidth]{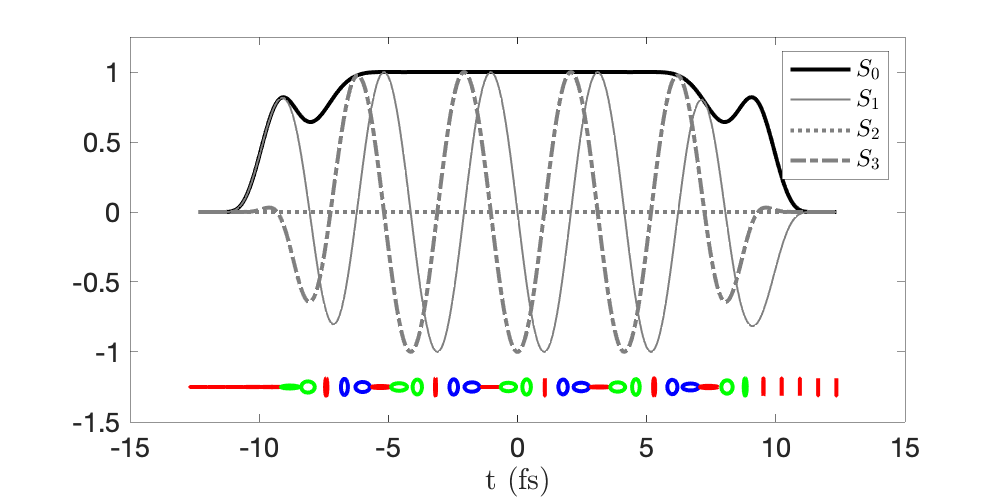}
   \caption{Temporal Stokes parameters in Eq.~(\ref{SHV}) for crossed planar undulators, same parameters as FIG.~\ref{fig:pulses}. The relative phase is set to $\phi=0$ (top), $\phi=\pi/4$ (middle), and $\phi=\pi/2$ (bottom). Also shown are the polarization ellipses along the pulse, with linear polarization red, $\mathbf{\hat e}_R$ polarization green, and $\mathbf{\hat e}_L$ polarization blue.}
   \label{fig:HV}
\end{figure}

\begin{figure}[h]
       \includegraphics*[width=.49\linewidth]{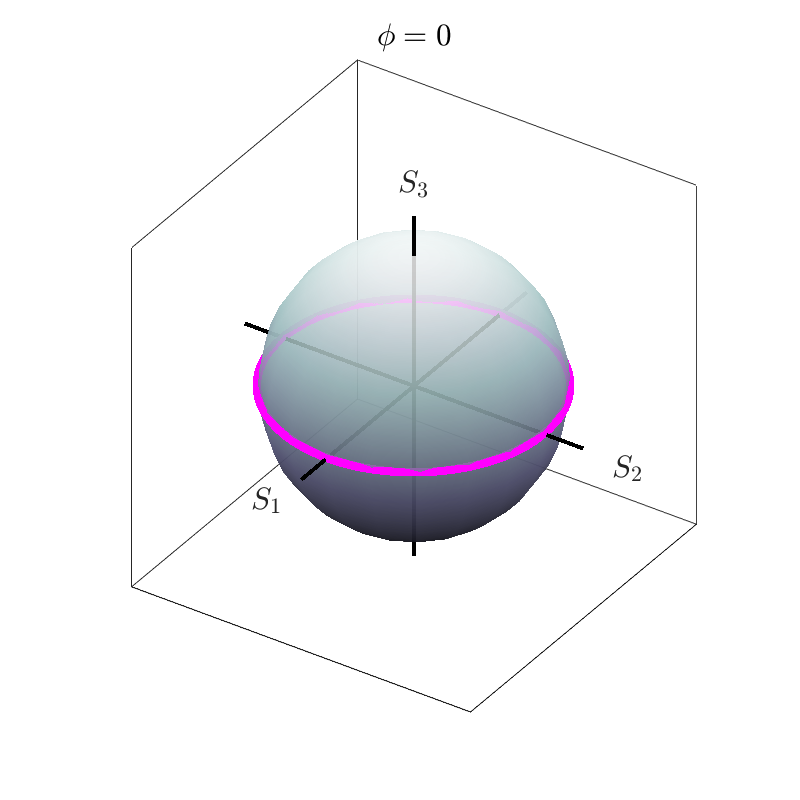}
         \includegraphics*[width=.49\linewidth]{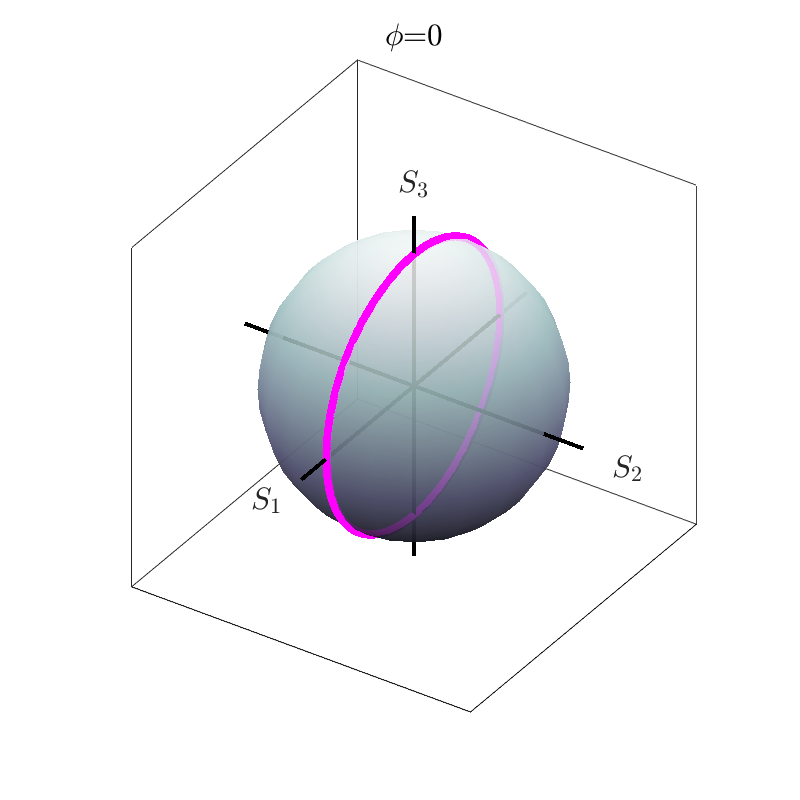}
        \includegraphics*[width=.49\linewidth]{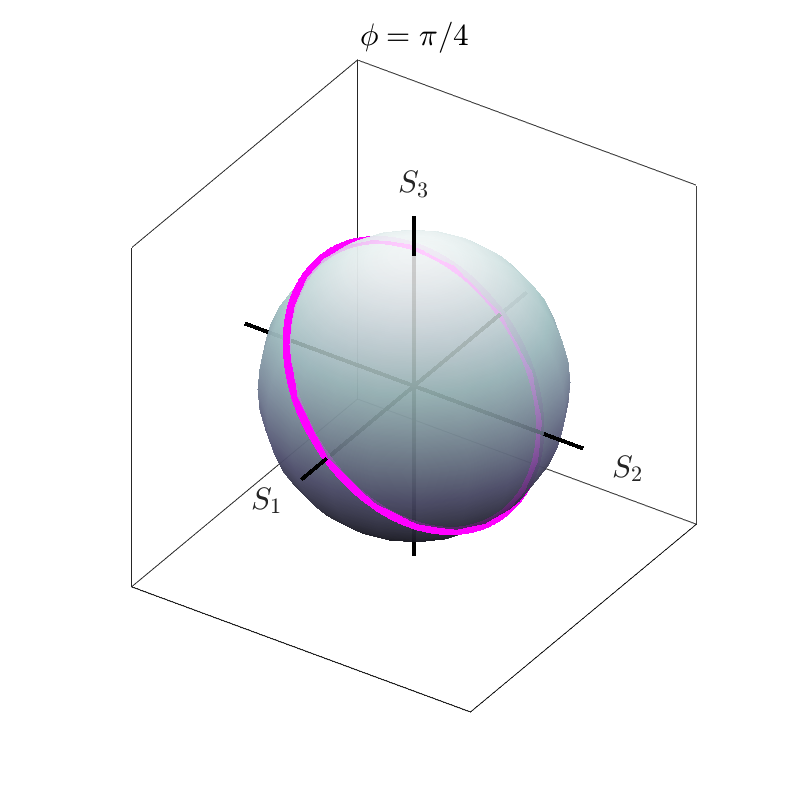}
          \includegraphics*[width=.49\linewidth]{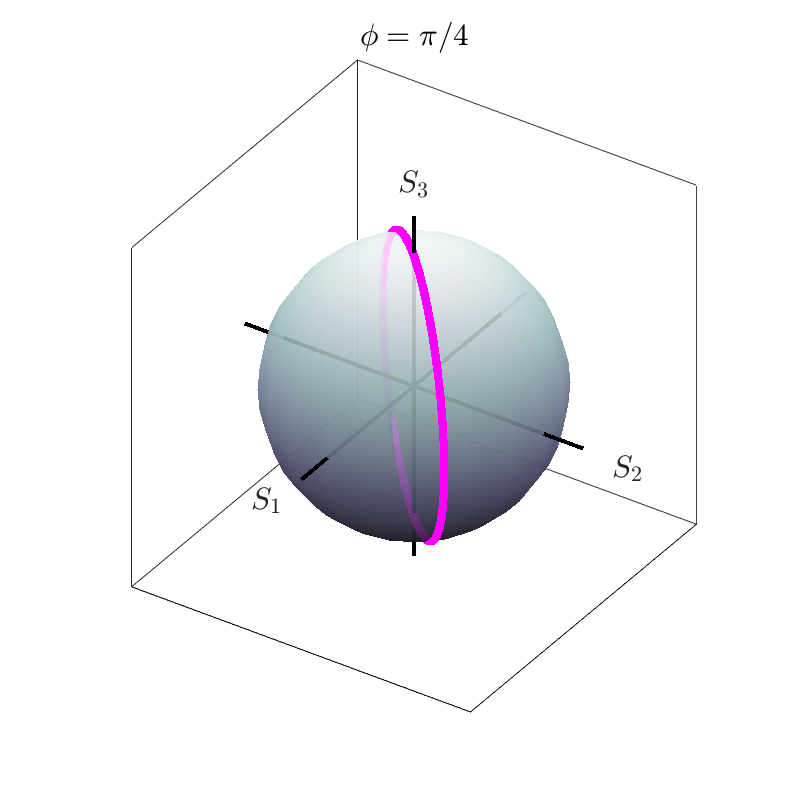}
         \includegraphics*[width=.49\linewidth]{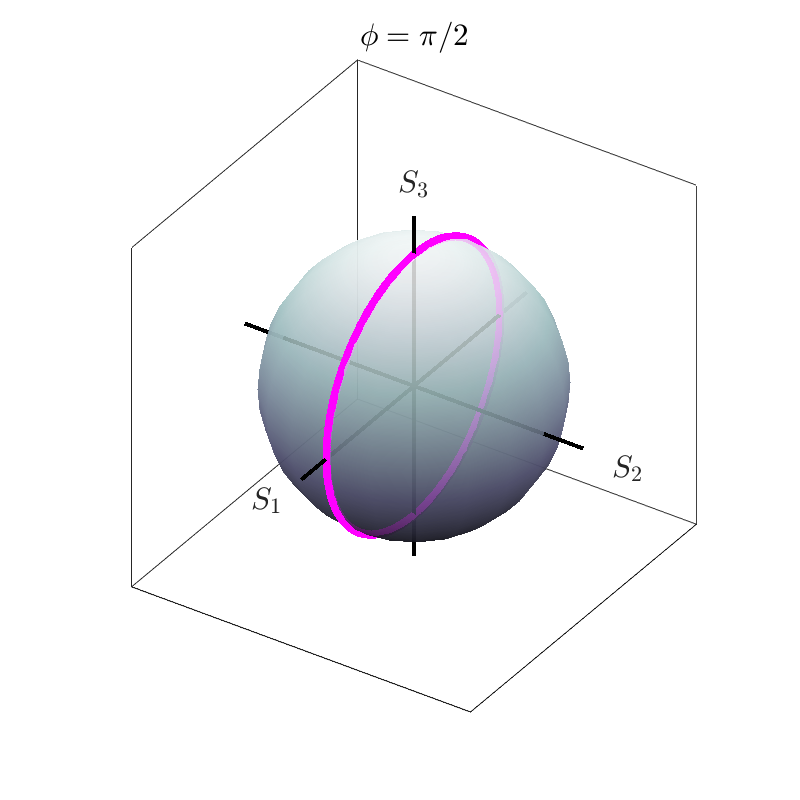}
           \includegraphics*[width=.49\linewidth]{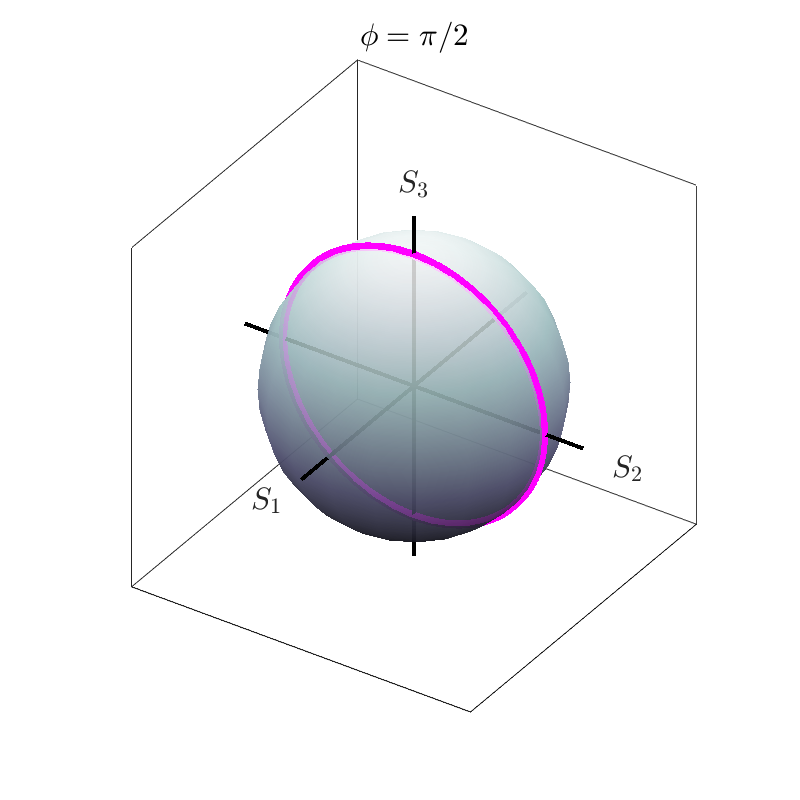}
   \caption{Distribution of Stokes parameters on the Poincar\'e sphere for two-color emission with crossed linear undulators (left) and crossed helical undulators (right). The time delay is $\Delta t=\pi/\Delta\omega$.}
   \label{fig:PS}
\end{figure}

The Stokes parameters for the crossed planar undulators are
\begin{equation}
\begin{aligned}
S_0&=1+\cos\left(\Delta\omega\, t\right)\cos\left(\frac{\Delta\omega \Delta t}{2}\right)\\
S_1&=-\sin\left(\Delta\omega\, t\right)\sin\left(\frac{\Delta\omega \Delta t}{2}\right)\\
S_2&=\cos\left(\phi\right)\left[\cos\left(\Delta\omega\, t\right)+\cos\left(\frac{\Delta\omega \Delta t}{2}\right)\right]\\
S_3&=-\sin\left(\phi\right)\left[\cos\left(\Delta\omega\, t\right)+\cos\left(\frac{\Delta\omega \Delta t}{2}\right)\right].
\end{aligned}
\end{equation}
We have defined a total phase shift between the two pulses as $\phi=\omega_0\Delta t-\psi$.

The intensity $S_0$ shows the temporal beat modulation with period $2\pi/\Delta\omega$. The amplitude of the modulation can be controlled by adjustment of the delay $\Delta t$ between the overlapping pulses. If the pulses are interleaved by setting the delay to half of the beat period, $\Delta t=\pi/\Delta\omega$, then the intensity modulation of the combined pulse vanishes (see FIG.~\ref{fig:pulses}). In this case the Stokes vector is,
\begin{equation}\label{SHV}
\mathbf{S}(t)=\begin{pmatrix} 
1 \\
-\sin\left(\Delta\omega\, t\right)\\
\cos\left(\phi\right)\cos\left(\Delta\omega\, t\right)\\
-\sin\left(\phi\right)\cos\left(\Delta\omega\, t\right)\\
\end{pmatrix}.
\end{equation}
Each of the polarization parameters $S_1,S_2,S_3$ has a time dependence with periodicity $2\pi/\Delta \omega$. The linear polarization parameter $S_1$ has a fixed amplitude, while the diagonal $S_2$ and circular polarization $S_3$ can be tuned in quadrature by the phase $\phi$.  Figure~\ref{fig:HV} shows the time-dependent Stokes parameters and polarization ellipses for different phase $\phi$ values. For $\phi=0$, the circular component vanishes $S_3=0$ and the polarization is linear at all times along the pulse, but evolves steadily between horizontal, diagonal, and vertical. In terms of the Poincar\'e sphere in FIG.~\ref{fig:PS}, the polarization state evolves strictly along the equator defined by the $(S_1,S_2)$ plane. For $\phi=\pi/2$, the diagonal component vanishes $S_2=0$, and the polarization evolves from linear to circular polarization purely in the $(S_1,S_3)$ plane. For $\phi=\pi/4$, the polarization involves all three parameters along a tilted plane, but is never purely circularly or diagonally polarized. 

The situation is similar if we instead use two orthogonally polarized helical undulators. In this case, the field polarization vectors are
\begin{equation}\label{vecs}
\mathbf{\hat e}_1=\mathbf{\hat e}_R=\frac{1}{\sqrt{2}}\begin{pmatrix} 
1 \\
-i
\end{pmatrix},\qquad \mathbf{\hat e}_2=\mathbf{\hat e}_L=\frac{1}{\sqrt{2}}\begin{pmatrix} 
1 \\
i
\end{pmatrix}
\end{equation}

The Stokes parameters are then modified to be
\begin{equation}
\begin{aligned}
S_0&=1+\cos\left(\Delta\omega\, t\right)\cos\left(\frac{\Delta\omega \Delta t}{2}\right)\\
S_1&=\cos\left(\phi\right)\left[\cos\left(\Delta\omega\, t\right)+\cos\left(\frac{\Delta\omega \Delta t}{2}\right)\right]\\
S_2&=-\sin\left(\phi\right)\left[\cos\left(\Delta\omega\, t\right)+\cos\left(\frac{\Delta\omega \Delta t}{2}\right)\right]\\
S_3&=\sin\left(\Delta\omega\, t\right)\sin\left(\frac{\Delta\omega \Delta t}{2}\right).
\end{aligned}
\end{equation}
\\
\\
Again interleaving the two temporal modulations with the delay $\Delta t=\pi/\Delta\omega$, for crossed helical undulators we obtain the Stokes vector
\begin{equation}\label{SLR}
\mathbf{S}(t)=\begin{pmatrix} 
1 \\
\cos\left(\phi\right)\cos\left(\Delta\omega\, t\right)\\
-\sin\left(\phi\right)\cos\left(\Delta\omega\, t\right)\\
\sin\left(\Delta\omega\, t\right)\\
\end{pmatrix}.
\end{equation}
Here, for $\phi=0$, the diagonal component $S_2$ vanishes, and the polarization evolves between circular and linear. This is similar to the $\phi=\pi/2$ case for cross planar undulators. At $\phi=\pi/2$ the $S_1$ component vanishes, and the polarization state evolves through pure diagonal and circular orientations in the $(S_2,S_3)$ plane. 

Overall, we see that the Stokes parameters all evolve with a period of $2\pi/\Delta\omega$. Therefore, at the shift $\Delta t=\pi/\Delta\omega$, the time it takes to change from one pure polarization state to another, (i.e., from $\mathbf{\hat e}_x\to\mathbf{\hat e}_R$) is 
\begin{equation}
\tau_{\mathbf{\hat e}_i\to\mathbf{\hat e}_j}=\frac{\pi}{2\Delta\omega}.
\end{equation}

For example, consider the case of a $\hbar\Delta\omega=1$~eV energy separation ($\Delta\lambda=1.24~\mu$m) between the two colors. This gives a temporal shift of $\Delta t=\pi/\Delta\omega=2$~fs. With crossed planar undulators, the polarization changes from pure linear polarization to pure circular polarization in $\tau_{\mathbf{\hat e}_x\to\mathbf{\hat e}_R}=1$~fs.

\begin{figure*}[t]
\includegraphics[scale=0.23]{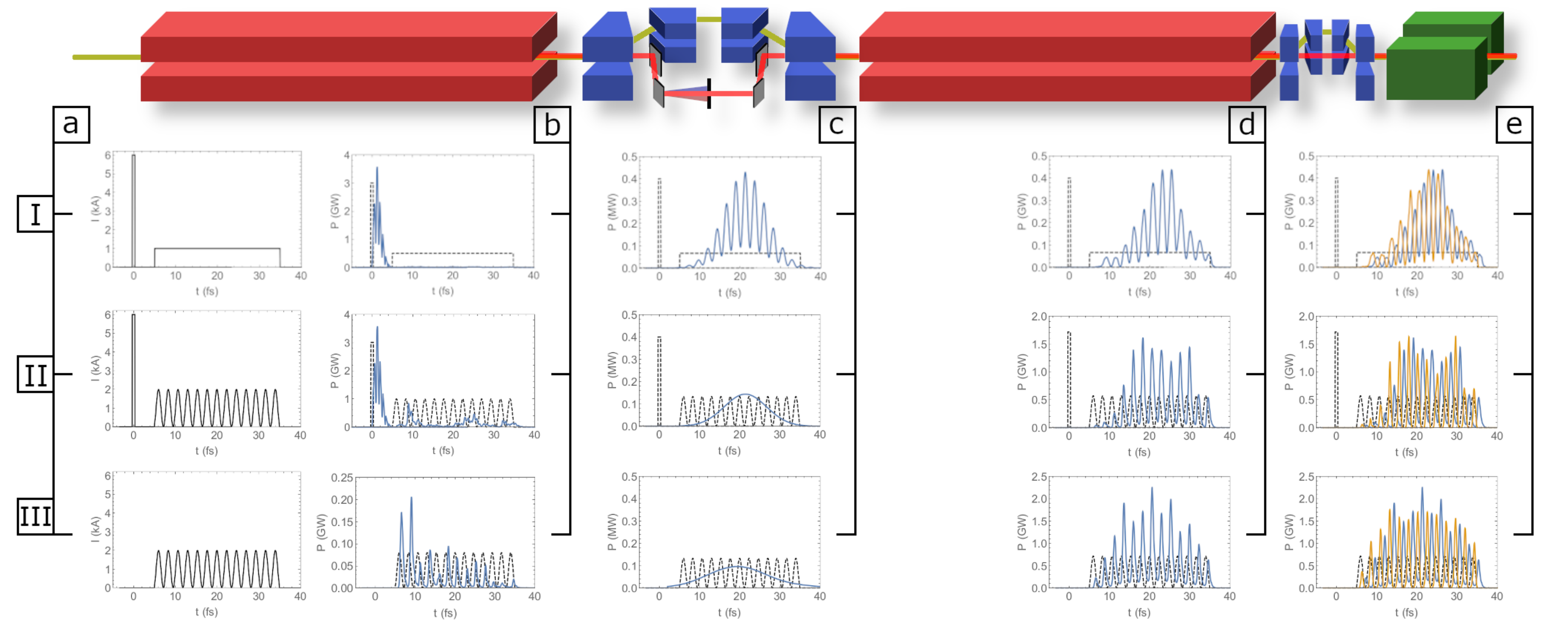}
\caption{Illustration of the methods considered for generation of time dependent polarization, showing the initial current profile (a), the SASE radiation generated before self seeding (b), the seed pulse after the self seeding monochromater (c), the $s$-polarization pulse (d), and the combined $s$ and $p$-polarization (blue, yellow) after the delay and crossed undulator (e), for two-color ESS (I), ESS with current modulation (II) and regular SS with current modulation (III). }
\label{layout}
\end{figure*}

\section{Time Varying Polarization: Methods}

Here we consider three configurations for generating x-ray pulses with time varying polarization at the fs time scale as described in the previous section.  Each configuration has trade-offs in the realms of available hardware, practical realization, stability and peak power.  For each case we consider intensity modulations of 700 nm (2.33 fs) and 1050 nm (3.5 fs), demonstrating the ability to achieve temporal structures in the visible and infrared.   

\subsection*{(I) Two Color Enhanced Self seeding}

Enhanced self seeding is a recently proposed method to improve intensity stability in the production of narrow band, highly coherent x-ray pulses.  Here, an ultrashort electron bunch with pulse length comparable to the FEL cooperation length generates a single SASE spike in the initial undulator section, reaching saturation.  The single spike nature of this pulse is characterized by a wide, coherent bandwidth.  Lasing in a nearby long, lower current electron bunch is suppressed in the SASE stage by its reduced gain length and undulator tapering, thus these electrons remain ``fresh".  For self-seeding, a narrow portion of the short spike spectrum is selected by a monochromater. This seed radiation is thus stretched, and then overlapped and amplified by the long fresh bunch in a subsequent undulator section.

It was shown in \cite{PhysRevLett.125.044801} that the filtered seed can exhibit a periodic intensity modulation by employing a monochromater capable of selecting two narrow regions of the single spike spectrum.  Selecting two phase-stable frequencies with separation $\Delta \omega$ will generate a beat wave modulation of the intensity with period $\lambda_m = 2\pi c/\Delta \omega$. The modulation is then preserved and amplified by the long bunch.  This two color enhanced self seeding provides a source of intensity stable, high power pulses with periodic intensity modulation at the femto-second time scale.  

We consider this as a candidate for arriving at the needed temporal intensity modulation for the time varying polarization scheme, as detailed in Figure 4, row I.  
In the seeded section, the intensity modulated two-color seed generates radiation in the $s$-polarization.  A small chicane then delays the electron beam without destroying the bunching in the regions of peak intensity, slipping the $s$-polarization pulse ahead by a half modulation period, $\lambda_m/2$.  A final crossed undulator section generates a similar pulse in the $p$-polarization.  

Figure 5 shows results from 3-D time dependent {\sc genesis} simulations considering idealized current profiles.  Here we show simulations of the two-color ESS method with frequency separation $\hbar \Delta \omega = 1.77$~eV and $\hbar \Delta \omega = 1.33$~eV corresponding to an intensity modulation of $\lambda_m = 700$~nm and $\lambda_m = 1050$~nm respectively.  Simulation parameters are provided in Table I, following the electron beam and undulator parameters available at the LCLS-II. 

The $s$-polarization pulse diffracts freely in the delay and $p$-polarization sections.  In order to approximately match the magnitude of the intensity from both polarizations, we propagate the simulated radiation profile through a lens such that the $s$ and $p$ polarizations are upstream and downstream of a waist respectively. With this in mind, the normalized Stokes parameters are plotted for the field with an approximate $\pi/2$ phase shift between the two polarizations.  Here we observe rotation of the polarization vector between linear and helical polarization states as shown in the simplified model, with linear polarization switching between $s$ and $p$ states of $\sim 90\%$ purity. 

A gentle undulator taper is introduced in the $s$-polarization undulator to suppress lasing in the troughs of the intensity modulation, however some bunching is generated in these regions.  This bunching can be significantly amplified by the delay between crossed undulators, spoiling the temporal structure of the electron beam bunching entering the $p$-polarization undulator.  This places a limitation on the achievable peak power, modulation period, and resonant wavelength. Furthermore, the radiation generated in the troughs will degrade the purity of the polarization states, since the troughs of one polarization overlap the peaks of the other. The achievable peak power and polarization purity may be improved by introducing a reverse taper in the undulator section following self-seeding to generate bunching only in the desired temporal regions and with significantly less growth of the energy spread \cite{PhysRevSTAB.16.110702}.

The practical realization of this method relies on the design of a two-color soft x-ray monochromator.  Whereas a comparable technique has been realized in hard x-ray self seeding \cite{PhysRevLett.113.254801}, a soft x-ray counterpart would require the fabrication of gratings with superimposed or alternating line densities.  This is currently being investigated \cite{2color}.  The ability to produce the short bunch-long bunch current profile in the LCLS-II copper linac was shown in simulations which are referenced in the supplementary material of \cite{PhysRevLett.125.044801}, and is a topic of active study.

\begin{figure}[h]
\includegraphics[scale=0.34]{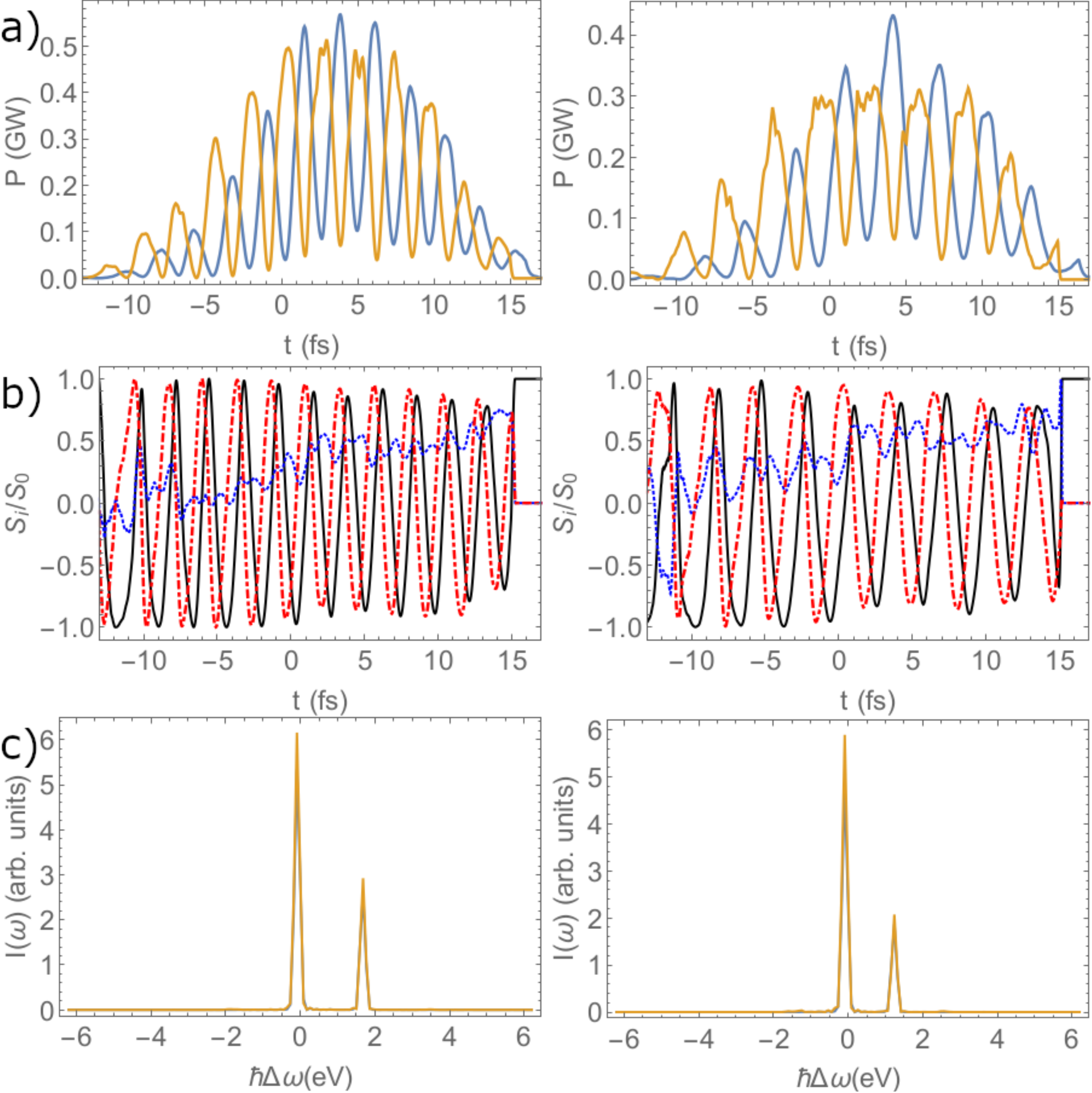}
\caption{Temporal modulation produced with two-color ESS. Left: 700 nm modulation Right: 1050 nm modulation. (a) Power in $s$-polarization (blue) and $p$-polarization (yellow). (b) Normalized Stokes parameters, $S_1/S_0$ (black, line), $S_2/S_0$ (blue, dotted), $S_3/S_0$ (red, dashed). (c) Spectral intensity in $s$-polarization (blue) and $p$-polarization (yellow).}
\label{layout}
\end{figure}

\subsection*{(II) Enhanced Self Seeding with Current Modulation}

Again taking advantage of the intensity stability offered by the ESS technique, in this method we can produce a temporal intensity modulation after the self seeding monochromator by modulating the current profile of the long bunch, as illustrated in Figure 4, row II.  Again a single SASE spike is generated by the short bunch in an initial undulator section, with lasing suppressed in the long bunch with undulator tapering.  This short pulse is then frequency filtered in a monochromator, this time selecting a single frequency. A current-modulated beam seeded by a single color coherent source will generate a series of coupled sidebands in the spectrum, spaced at the current modulation frequency.  This results in a series of mode-locked pulses in the time domain with $s$-polarization, as described in \cite{Kur_2011}.  As before, a small chicane then delays the electron beam, preserving the bunching in the current spikes and slipping the $s$-polarization pulse ahead by a half modulation period.  A final crossed undulator section generates a similar pulse in the $p$-polarization. 

\begin{figure}
\includegraphics[scale=0.37]{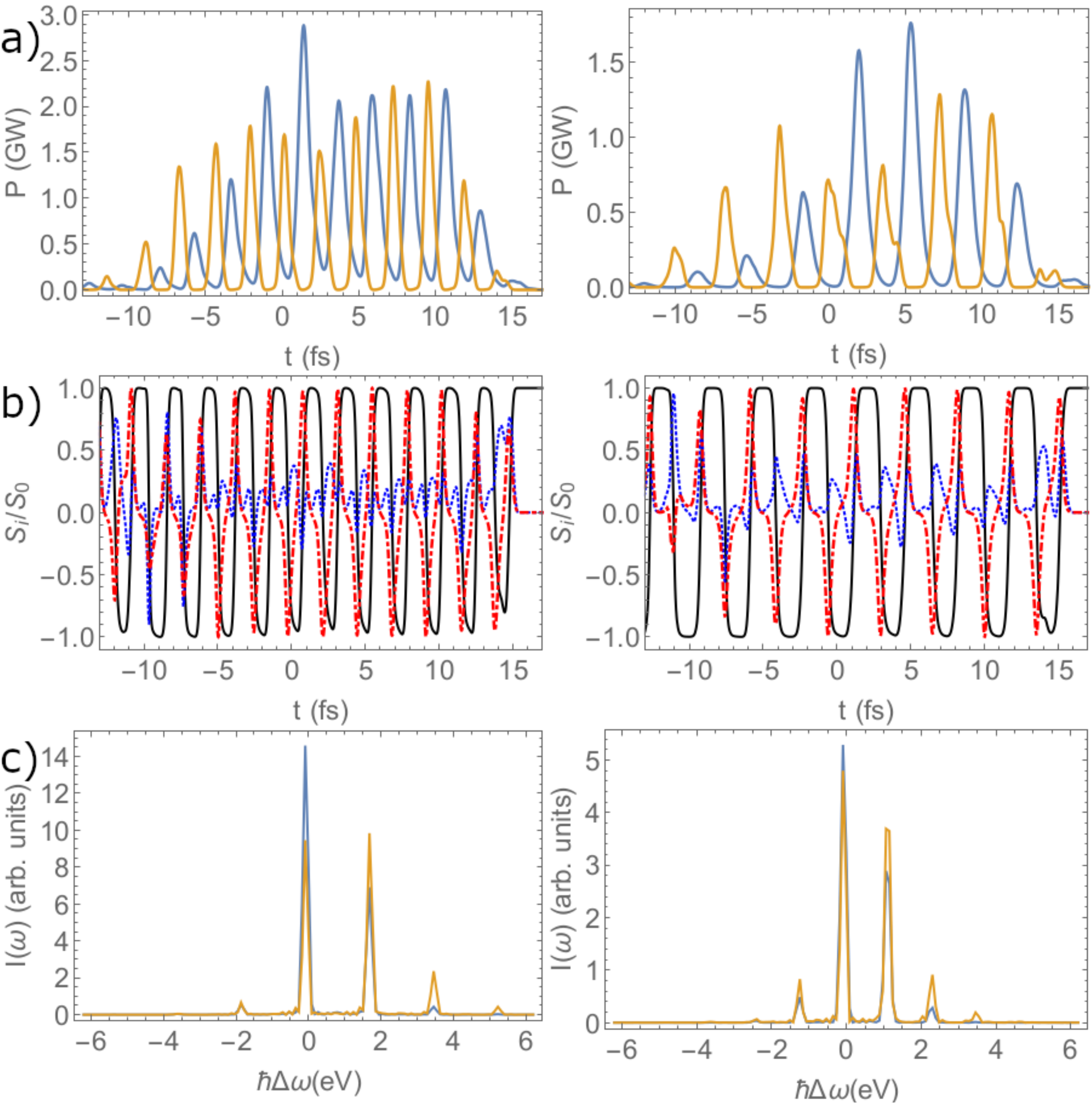}
\caption{Temporal modulation produced with single color enhanced self-seeding plus current modulation.  Left: 700 nm modulation Right: 1050 nm modulation. (Top) Power in $s$-polarization (blue) and $p$-polarization (yellow). (middle) Normalized Stokes parameters, $S_1/S_0$ (black, line), $S_2/S_0$ (blue, dotted), $S_3/S_0$ (red, dashed).   (Bottom) On-axis spectral intensity in $s$-polarization (blue) and $p$-polarization (yellow).}
\label{layout}
\end{figure}

Figure 6 shows results from 3-D time dependent {\sc genesis} simulations considering flattop electron beam current profiles superimposed with modulations of $\lambda_m = 700$~nm and $\lambda_m = 1050$~nm.  Figure 6 also shows the normalized Stokes parameters for the field with a $\pi/2$ phase shift between the two polarizations.  Here we observe rotation of the polarization vector between linear and helical polarization states with linear polarization switching between $s$ and $p$ states of nearly $100\%$ purity.  The achievable peak power is increased compared to the two color ESS method due to the increased peak current. To optimize this setup, both the $s$ and $p$ undulator sections are longer than in the first method (see Table 1), which reduces the required delay.  In this example we take the radiation closer to saturation, which lead to the growth of additional spectral modes and produces sharp intensity spikes \cite{BONIFACIO1988369,PhysRevA.42.4120}. This has the effect of reducing the overlap between polarizations and making the transition between polarization states more abrupt. We note that, while the use of ESS may provide shot-to-shot stability in these temporal structures, the realization of the requisite short beam plus time-modulated current profile in the long beam may prove difficult.

\begin{table}[H]
\caption{\label{parameters} Genesis simulation parameters}
\begin{ruledtabular}
\begin{tabular}{lc}
\textrm{Parameter}&
\textrm{Value}\\
\colrule
e-beam energy (GeV) & 4 \\
e-beam emittance $\epsilon_{x,y}$ ($\mu$m) & 0.4 \\
e-beam beta function $\beta_{x,y}$ (m) & 12 \\
undulator period $\lambda_w$ (cm) & 3.9\\
undulator section length (m) & 3.4\\
break section length (m) & 0.975\\
fundamental photon energy $\hbar \omega_0$ (eV) & 620\\
Self seeding (S.S.) chicane $R_{56}$ ($\mu$m) & 750\\
S.S. filter RMS (meV) & 85 \\
\end{tabular}
\begin{tabular}{lcc}
\textrm{Method I}&
\textrm{700 nm mod.}&
\textrm{1050 nm mod.}\\
\colrule
short e-beam I (kA) & 6 & 6 \\
short e-beam FWHM (fs) & 0.67 & 0.67 \\
short e-beam $\sigma_E$ (MeV) & 3 & 3\\
long e-beam I (kA) & 1 & 1 \\
long e-beam FWHM (fs) & 30 & 30 \\
long e-beam $\sigma_E$ (MeV) & 0.5 & 0.5\\
pre S.S. $N_{sec}$ & 9 & 9\\
pre S.S. K & 2.3189-2.2818 & 2.3189-2.2818\\
s-polarization $N_{sec}$ & 6 & 6\\
s-polarization K & 2.2982-2.2935& 2.3011-2.2952\\
chicane $R_{56}$ ($\mu$m) & 0.36 & 0.6 \\
p-polarization $N_w$ & 40 & 32\\
p-polarization K & 2.2937 & 2.2937 \\
\end{tabular}
\begin{tabular}{lcc}
\textrm{Method II}&
\textrm{700 nm mod.}&
\textrm{1050 nm mod.}\\
\colrule
short e-beam I (kA) & 6 & 6 \\
short e-beam FWHM (fs) & 0.67 & 0.67 \\
short e-beam $\sigma_E$ (MeV) & 3 & 3\\
long e-beam peak I (kA) & 2 & 2 \\
long e-beam FWHM (fs) & 30 & 30 \\
long e-beam $\sigma_E$ (MeV) & 1 & 1\\
pre S.S. $N_{sec}$ & 9 & 9\\
pre S.S. K & 2.3189-2.2818 & 2.3189-2.2818\\
s-polarization $N_{sec}$ & 7 & 6\\
s-polarization K & 2.3011-2.2941& 2.2982-2.2935\\
chicane $R_{56}$ ($\mu$m) & 0.2 & 0.5 \\
p-polarization $N_w$ & 55 & 50\\
p-polarization K & 2.2912 & 2.2912 \\
\end{tabular}
\begin{tabular}{lcc}
\textrm{Method III}&
\textrm{700 nm mod.}&
\textrm{1050 nm mod.}\\
\colrule
long e-beam peak I (kA) & 2 & 2 \\
long e-beam FWHM (fs) & 30 & 30 \\
long e-beam $\sigma_E$ (MeV) & 1 & 1\\
pre S.S. $N_{sec}$ & 6 & 6\\
pre S.S. K & 2.2937 & 2.2937\\
s-polarization $N_{sec}$ & 6 & 5\\
s-polarization K & 2.2937& 2.2937\\
chicane $R_{56}$ ($\mu$m) & 0.3 & 0.7 \\
p-polarization $N_w$ & 45 & 25\\
p-polarization K & 2.2967 & 2.2937 \\
\end{tabular}
\end{ruledtabular}
\end{table}

\subsection*{(III) Regular Self Seeding with Current Modulation}

The previous method can be simplified in practice by using the current modulated long bunch to generate its own seed in a standard self seeding configuration, as shown in Figure 4, row III.  In this case, a SASE pulse is generated by the long current-modulated bunch in an initial undulator section.  Due to the large number of cooperation lengths in the beam this pulse is stochastic in nature, exhibiting significant fluctuations in temporal and spectral intensity. This pulse is then frequency filtered in a monochromator, and the seed is amplified in the downstream section by the same current-modulated beam to produce a temporally-modulated pulse.



Figure 7 shows results from simulations with a $\pi/2$ phase shift between the two polarizations.  Here we observe similar rotation of the polarization vector between linear and helical polarization states with linear polarization switching between $s$ and $p$ states of nearly $100\%$ purity, however with a reduction of the interleaving helical states due to the even more pronounced sharpness of the modulation.  This method of conventional SS will likely exhibit greater shot-to-shot fluctuations than the previous schemes that use ESS, but it is simpler to setup and implement. Start to end simulations of this method, including the production of the modulated current profile, are presented in the next section.  


\begin{figure}
\includegraphics[scale=0.35]{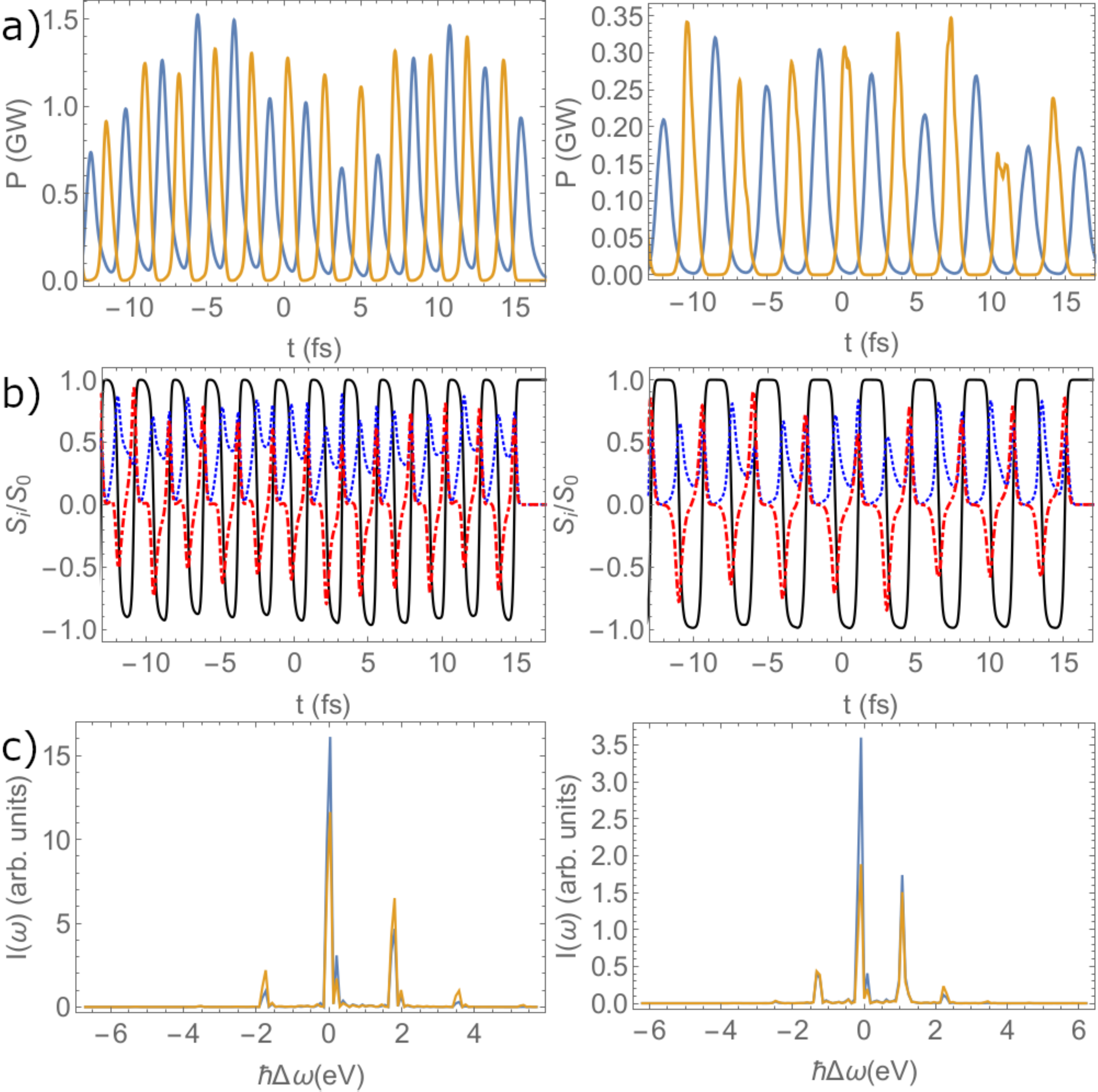}
\caption{Temporal modulation produced with single color regular self-seeding plus current modulation. Left: 700 nm modulation Right: 1050 nm modulation. (Top) Power in $s$-polarization (blue) and $p$-polarization (yellow). (middle) Normalized Stokes parameters, $S_1/S_0$ (black, line), $S_2/S_0$ (blue, dotted), $S_3/S_0$ (red, dashed).   (Bottom) On-axis spectral intensity in $s$-polarization (blue) and $p$-polarization (yellow).}
\label{layout}
\end{figure}

\section{Start to End simulations}

In order to generate a fs-scale current modulation we consider seeding the micro-bunching instability at the laser heater with a THz beatwave-modulated laser pulse, as was demonstrated in \cite{PhysRevLett.115.214801}.  Using the chirped-pulse beating technique, a THz scale intensity-modulated laser pulse is generated by the 1 $\mu$m wavelength laser heater laser \cite{Weling:96}.  Overlapping this pulse with the electron beam in the laser heater undulator will generate a corresponding energy modulation.  Bunch compression in the downstream electron beam transport will convert this energy modulation to density modulation, with further amplification of the energy modulation provided by longitudinal space charge impedance \cite{PhysRevLett.111.034803}.  Bunch compression also serves to reduce the modulation period by the compression factor of the accelerator, reaching the desired femtosecond level.  

Figure 8 shows the laser heater laser intensity profile and energy modulation at the laser heater chicane exit from {\sc elegant} simulations \cite{osti_761286}, with an initial beam distribution generated from {\sc impact} simulations of the LCLS-II gun \cite{QIANG2000434,PhysRevSTAB.9.044204}.  Simulation parameters are given in Table II. 

\begin{table}
\caption{\label{parameters} Parameters for start-to-end simulations}
\begin{ruledtabular}
\begin{tabular}{lcdr}
\textrm{Elegant parameters}&
\textrm{Value}\\
\colrule
laser heater laser peak power (MW) & 50 \\
laser heater laser waist ($\mu$m) & 100 \\
laser heater laser wavelength ($\mu$m) & 1.03\\
laser heater modulation period ($\mu$m) & 75 \\
e-beam charge (pC) & 100 \\
e-beam energy at laser heater (MeV) & 100 \\
e-beam energy at undulator (GeV) & 4 \\
e-beam emittance $\epsilon_{x,y}$ (mm-mrad) & 0.39,0.37\\
e-beam beta function $\beta_{x,y}$ (m) & 12,12\\
Compressed current modulation period ($\mu$m) & 0.92 \\
X-leap undulator period (cm) & 35\\
X-leap undulator length (m) & 2.1\\
X-leap undulator strength (rms) K & 17.92\\
X-leap chicane $R_{56}$ (mm) & 1.2\\
\end{tabular}
\begin{tabular}{lcc}
\textrm{{\sc genesis} parameters}&
\textrm{No X-Leap}&
\textrm{w/ X-Leap}\\
\colrule
e-beam peak I (kA) & 1.75 & 3 \\
e-beam modulation amplitude (kA) & 1 & 2.75 \\
e-beam $\sigma_E$ (MeV) & 0.4 & 1\\
pre S.S. $N_{sec}$ & 5 & 5\\
pre S.S. K (rms) & 2.2937 & 2.2937\\
$s$-polarization $N_{sec}$ & 6 & 5\\
$s$-polarization K (rms) & 2.2937& 2.2937\\
chicane $R_{56}$ ($\mu$m) & 0.4 & 0.5 \\
p-polarization $N_w$ & 40 & 45\\
p-polarization K (rms) & 2.2937 & 2.2937 \\
\end{tabular}
\end{ruledtabular}
\end{table}

\begin{figure}
\includegraphics[scale=0.3]{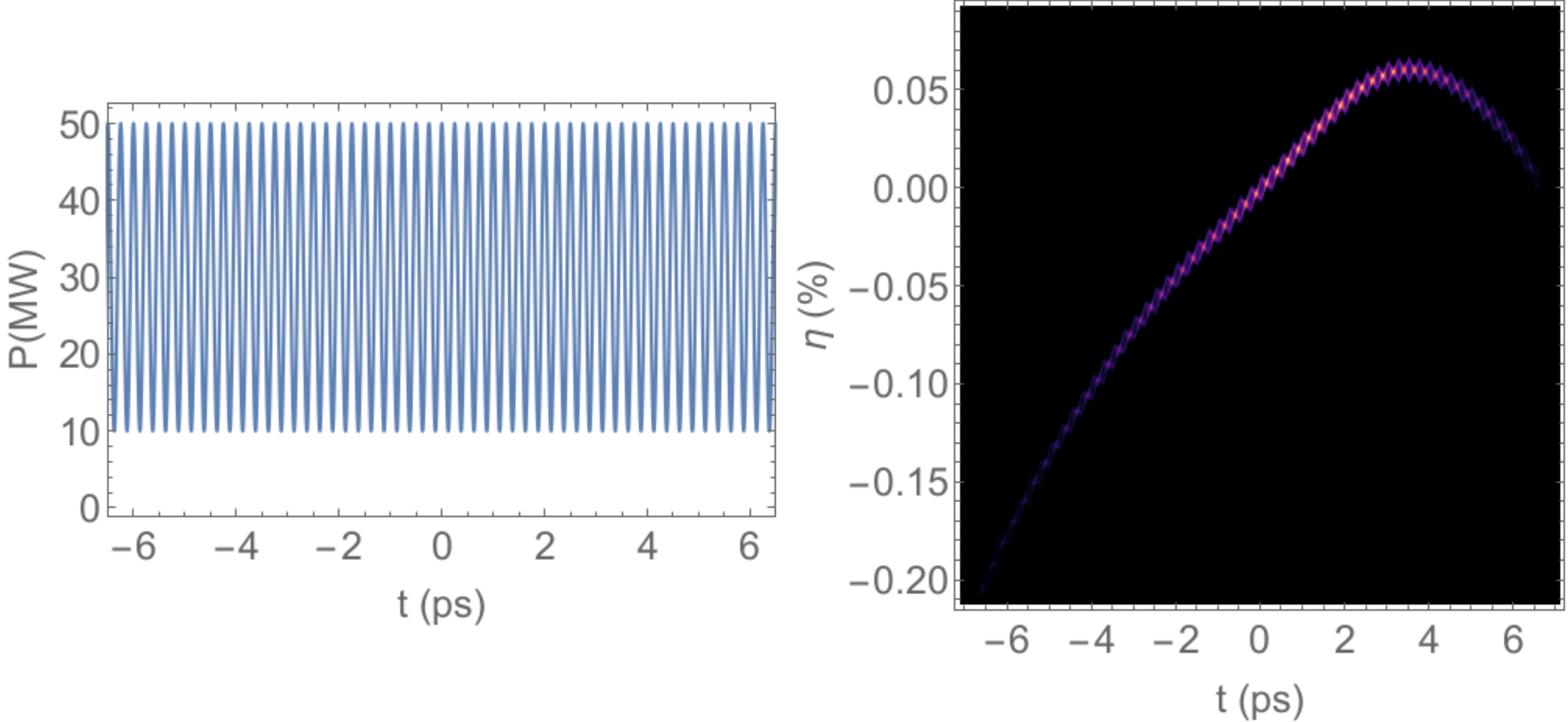}
\caption{(left) Laser heater modulation. (right) longitudinal phase space at laser heater exit}
\label{layout}
\end{figure}

The $\lambda_m = 920$~nm current modulation at the undulator entrance is shown in Figure 9. To further increase the current spikes, we pass this current modulated beam through a high impedance LCLS-II X-Leap wiggler magnetic wiggler, which will produce an additional sinusoidal energy modulation driven by the wiggler's short range CSR wake, as was demonstrated in \cite{Zhang_2020}.  Including an additional $R_{56}$ from a magnetic chicane will then produce a sharper current modulation.  This self modulation process can be simulated approximately in 1-D as described in \cite{PhysRevLett.123.214801}.  
Parameters are given in Table II, with the final longitudinal phase space shown in Figure 9.

Figure 9 also shows results from {\sc genesis} simulations with the final start-to-end beam. 
All simulations include the effects of the undulator chamber resisitive wall wakefields.  
From the normalized Stokes parameters 
we observe rotation of the polarization vector between linear and helical polarization states with linear polarization switching between $s$ and $p$ states of nearly $100\%$ purity.  Non-linearities in the longitudinal phase space cause a lengthening of the modulation period at the head and tail of the beam degrading the polarization switching.  Lasing in the head and tail could perhaps be suppressed by additional shaping of the laser heater laser. 

\begin{figure}[H]
\includegraphics[scale=0.35]{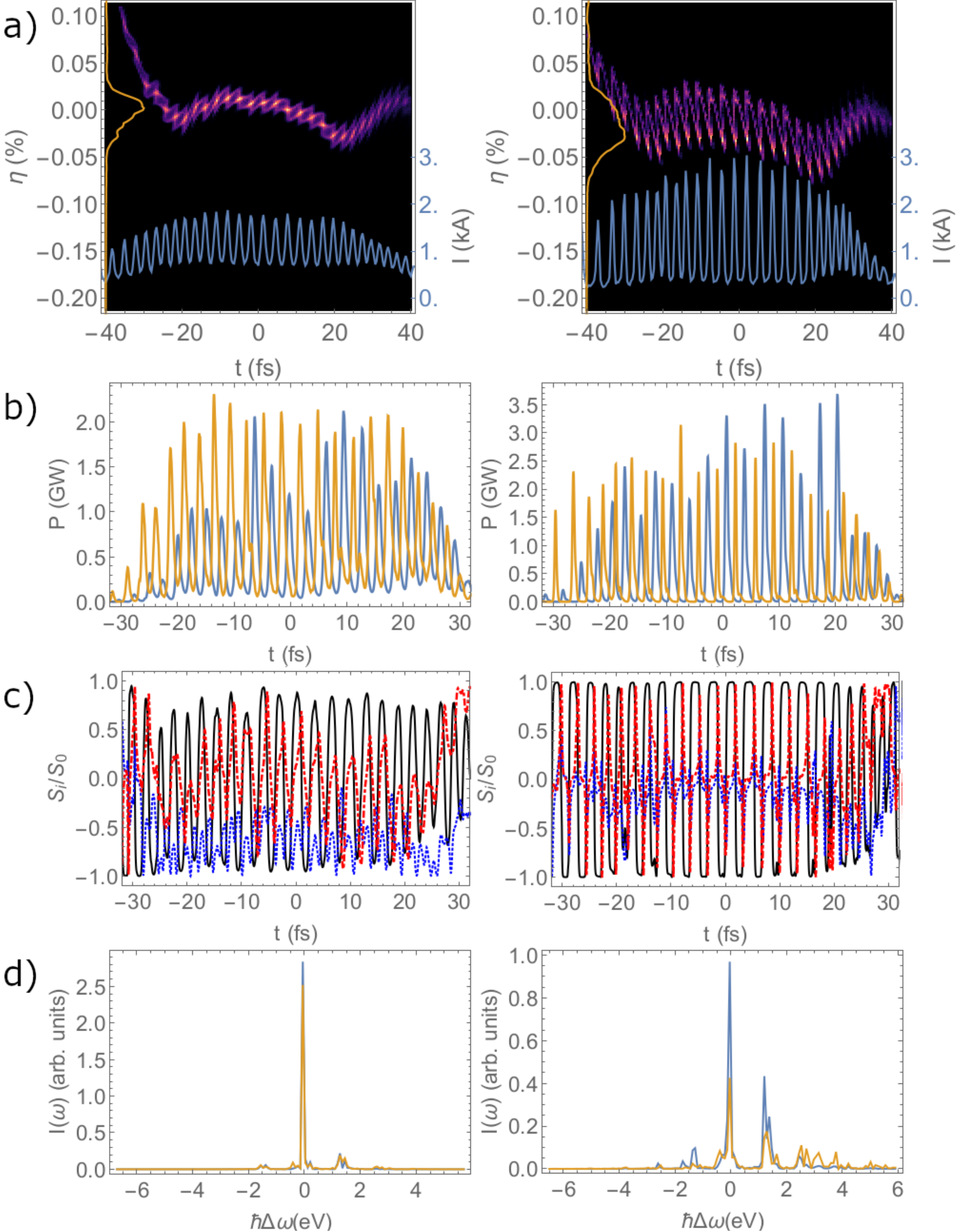}
\caption{Left: No X-Leap undulator modulation Right: With X-Leap undulator modulation and $R_{56}$. (a) Longitudinal phase space at undulator entrance with current profile included (blue). (b) Power in $s$-polarization (blue) and $p$-polarization (yellow). (c) Normalized Stokes parameters, $S_1/S_0$ (black, line),$S_2/S_0$ (blue, dotted), $S_3/S_0$ (red, dashed).   (d) On-axis spectral intensity in $s$-polarization (blue) and $p$-polarization (yellow).}
\label{layout}
\end{figure}

\section{conclusion}
The ability to generate x-ray pulses with femtosecond level polarization switching would provide a useful tool to the scientific community. Simulations of the three aforementioned methods demonstrate tunability of this temporal polarization modulation from the visible to infrared.  Each method exhibits benefits of stability, peak power, or realizability with existing hardware at LCLS-II.  Further investigation of the underlying scheme could lead to increases in the achievable peak power, especially at longer modulation period.  This could include implementing a reverse tapering scheme to better control the micro-bunching and energy spread in the $s$ and $p$ polarization stages. 

\begin{acknowledgments}
The authors would like to thank A. Marinelli for useful discussions. This work was
supported by U.S. Department of Energy Contract No. DE-AC02-76SF00515 and award no. 2017-SLAC-100382.
RNC acknowledges additional support from DOE Office of Science BES under Field Work Proposal 100498.
\end{acknowledgments}


\bibliographystyle{unsrt}
\bibliography{refs,TimeDependentPolarization_PRABNotes}
\end{document}